\documentclass[12pt]{article}
\pdfoutput=1

\usepackage{amsmath,amssymb,amscd}
\usepackage{listings}
\usepackage{caption}
\usepackage{dsfont}
\usepackage{slashed}
\usepackage{color}

\usepackage[pdftex]{graphicx}
\usepackage{epstopdf}
\usepackage{subfigure}
\usepackage{epsfig}
\usepackage{listings}
\usepackage{caption}
\usepackage{cite}

\usepackage{multirow}

\newcommand{\bea}{\begin{align}}
\newcommand{\eea}{\end{align}}
\newcommand{\beq}{\begin{equation}}
\newcommand{\eeq}{\end{equation}}
\newcommand{\nbea}{\begin{align*}}
\newcommand{\neea}{\end{align*}}
\newcommand{\nbeq}{\begin{equation*}}
\newcommand{\neeq}{\end{equation*}}
\newcommand{\beaa}{\begin{eqnarray}}  
\newcommand{\eeaa}{\end{eqnarray}}

\numberwithin{equation}{section}

\begin{document}

\begin{titlepage}

\pagestyle{empty}

\baselineskip=21pt
\rightline{KCL-PH-TH/2014-15, LCTS/2014-14, CERN-PH-TH/2014-061}

\vskip 1in

\begin{center}

{\large {\bf Complete Higgs Sector Constraints on Dimension-6 Operators}}

\end{center}

\begin{center}

\vskip 0.6in

 {\bf John~Ellis}$^{1,2}$,
 {\bf Ver\'onica Sanz}$^{3}$
and {\bf Tevong~You}$^{1}$

\vskip 0.4in

{\small {\it

$^1${Theoretical Particle Physics and Cosmology Group, Physics Department, \\
King's College London, London WC2R 2LS, UK}\\
$^2${TH Division, Physics Department, CERN, CH-1211 Geneva 23, Switzerland}\\
$^3${Department of Physics and Astronomy, University of Sussex, Brighton BN1 9QH, UK}\\
}}

\vskip 0.5in

{\bf Abstract}

\end{center}

\baselineskip=18pt 

\vskip 0.2in

\noindent

Constraints on the full set of Standard Model dimension-6 operators have previously used 
triple-gauge couplings to complement the constraints obtainable from Higgs signal strengths. 
Here we extend previous analyses of the Higgs sector constraints by including information
from the associated production of Higgs and massive vector bosons (H+V production), which
excludes a direction of limited sensitivity allowed by partial cancellations in the triple-gauge sector measured at LEP.
Kinematic distributions in H+V production provide improved sensitivity to dimension-6 operators, 
as we illustrate here with simulations of the invariant mass and $p_T$ distributions
measured by D0 and ATLAS, respectively.
We provide bounds from a global fit to a complete set of CP-conserving operators affecting Higgs physics.



\vfill

\leftline{April 2014}

\end{titlepage}

\newpage


\section{Introduction}

The investigation of the properties of the recently-discovered Higgs boson~\cite{Eureka} proceeded initially
by characterizing its signal strength relative to the Standard Model (SM) expectation~\cite{strengths}, 
with many studies refining this picture to constrain deviations in the Higgs couplings 
under various assumptions~\cite{fits}. Although the signal strengths and pattern of couplings provided 
some information about the spin and parity of the Higgs boson~\cite{primafacie}, it was through the use of 
differential kinematic distributions that different Lorentz structures could be probed most thoroughly~\cite{spinparity}.
The evidence now indicates convincingly~\cite{spin0,D0spin} that we are dealing with a spin-zero,
positive-parity particle, as expected for the Higgs boson responsible for electroweak symmetry breaking.

Moreover, there is no significant indication of any deviation of the dimension-4 couplings
of this particle from those expected in the SM. Studies of these couplings continue, and
are being supplemented by searches for anomalous couplings that could arise from new physics 
in the electroweak sector. If this new physics is decoupled at some heavy scale, 
then the effects of these interactions are cohesively captured by supplementing the SM Lagrangian
with higher-dimensional operators involving multiple fields and/or derivative interactions in
an effective field theory (EFT) framework\footnote{For a recent short review, see \cite{EFTreview}.}~\cite{eft}. 

Constraints on these operators have been placed for subsets of operators~\cite{dim6subset,eduard}
and in full global fits both before~\cite{hanandskiba} and after~\cite{dim6global, pomarolriva} the Higgs discovery\footnote{Ref.~\cite{pomarolriva} in particular includes a full set of operators in the EWPT sector.}. Many strong constraints come 
from electroweak precision tests (EWPT)~\cite{ewpt} at LEP, and from triple-gauge coupling (TGC) ~\cite{dim6global,tgc}
measurements at LEP and the LHC. 
In the case of Higgs observables, aside from operators contributing to couplings that are
absent at tree-level in the SM, only weaker limits are available so far. Some combinations 
of these operators enter into EWPT and TGC, but the presence of a poorly constrained direction~\cite{leshouches} in measurements of the 
latter means that constraints on dimension-6 operators from Higgs physics are complementary
and not redundant within the EFT framework. 
Constraints from EWPT on operators that contribute at loop level rely on assuming no unnatural cancellations~\cite{RGEbounds}, with unambiguous bounds being far weaker~\cite{dawson}. Thus, it is desirable to refine as much as possible the analysis of the Higgs sector~\cite{precisionhiggs}.

We illustrate here the power of associated $H + V$ production and its differential kinematic distributions
to constrain CP-conserving dimension-6 operators within the EFT framework. In particular, we note that
the distribution of the $H + V$ invariant mass, $m_{VH}$, 
measured by D0~\cite{D0comb} and the vector-boson transverse momentum, $p^V_T$, distribution
measured by ATLAS~\cite{atlasptv} in the associated production channel $V + H \to V \bar{b}b$ 
have very low backgrounds in the higher mass and $p_T$ bins, respectively,
where higher-dimension operators would contribute. These searches are, therefore,
ideal for constraining the boosted signature of new physics that could arise from dimension-6 operators,
despite the large uncertainties in the total signal strength~\cite{ESY2}. Moreover,
we find that the inclusion of associated production at D0 and ATLAS
removes certain degeneracies in a complete fit to the full set of operators affecting Higgs physics. 

In the following Section we introduce the CP-even dimension-6 operators that affect Higgs physics. 
In Section \ref{D0} we constrain one operator using the $m_{VH}$ distribution of $VH\to V\bar{b}b$ 
in the $V \to$ 0-, 1- and 2-lepton sub-channels used in the D0 search, 
quantifying the improvement obtained by using  differential information,
and we do the same using the ATLAS $p^V_T$ distribution in Section \ref{ATLAS}. 
In Section \ref{global} we combine these channels and make a multi-parameter fit to obtain
global constraints from the Higgs sector. Section \ref{conx} summarizes our conclusions.
Details of the analysis implementations for D0 and ATLAS can be found in the Appendices.

\section{Dimension-6 Operators in the Higgs Sector}
\label{definitions}

In the basis of~\cite{silh}, the CP-even dimension-6 Lagrangian involving Higgs doublets
may be written as 
\beaa
{\cal L} &\supset &    \frac{\bar c_{H}}{2 v^2} \partial^\mu\big[\Phi^\dag \Phi\big] \partial_\mu \big[ \Phi^\dagger \Phi \big] +  \frac{g'^2\ \bar c_{\gamma}}{m_W^2} \Phi^\dag \Phi B_{\mu\nu} B^{\mu\nu}
   + \frac{g_s^2\ \bar  c_{g}}{m_W^2} \Phi^\dag \Phi G_{\mu\nu}^a G_a^{\mu\nu} \nonumber\\
  &+& \frac{2 i g\ \bar c_{HW}}{m_W^2} \big[D^\mu \Phi^\dag T_{2k} D^\nu \Phi\big] W_{\mu \nu}^k 
  + \frac{i g'\ \bar c_{HB}}{m_W^2}  \big[D^\mu \Phi^\dag D^\nu \Phi\big] B_{\mu \nu} \nonumber\ \\
  & +  &\frac{i g\ \bar c_{W}}{m_W^2} \big[ \Phi^\dag T_{2k} \overleftrightarrow{D}^\mu \Phi \big]  D^\nu  W_{\mu \nu}^k 
  +\frac{i g'\ \bar c_{B}}{2 m_W^2} \big[\Phi^\dag \overleftrightarrow{D}^\mu \Phi \big] \partial^\nu  B_{\mu \nu}  \nonumber\ \\
  & + & \frac{\bar c_{t}}{v^2} y_t    \Phi^\dag \Phi\ \Phi^\dag\cdot{\bar Q}_L t_R + \frac{\bar c_{b}}{v^2} y_b     \Phi^\dag \Phi\ \Phi \cdot {\bar Q}_L b_R + \frac{\bar c_{\tau}}{v^2} y_\tau\ \Phi^\dag \Phi\ \Phi \cdot {\bar L}_L \tau_R 	\, .
\eeaa
%
%
%
We note that $\bar{c}_T$ corresponds to the $\hat{T}$ parameter,
which is constrained at the per-mille level by EWPT,
and $\bar{c}_6$ only affects the Higgs self-coupling, so we drop these from our analysis. 
The linear combination $\bar{c}_W + \bar{c}_B$ is related to the $\hat{S}$ parameter,
which is also bounded at the per-mille level, so we set $\bar c_B = -\bar c_W$.
The independent set of parameters affecting Higgs physics is thereby reduced to
\beaa
\bar{c}_i \equiv \{\bar{c}_H, \bar{c}_{t,b,\tau}, \bar c_W, \bar{c}_{HW}, \bar{c}_{HB}, \bar{c}_\gamma, \bar{c}_g  \} 	\, . 
\label{eq:fullset}
\eeaa
The other dimension-6 operators enter either in EWPT or TGC observables, but do not affect the Higgs sector. 
For an analysis of the above operators and TGCs, see Ref.~\cite{dim6global,tgc}.

A more phenomenological and experimentally transparent approach is often used in the form of an effective
Lagrangian  with anomalous Higgs couplings. Experimental bounds expressed 
in terms of anomalous couplings may then be related to other more theoretically-motivated effective theories or models,
which has proven to be a useful approach for EWPT and TGCs. For example,
following Ref.~\cite{benj}, the relevant subset of the Higgs anomalous couplings in the mass basis and unitary gauge includes 
\beaa
 {\cal L} \supset &  - &  \frac{1}{4} g_{HZZ}^{(1)} Z_{\mu\nu} Z^{\mu\nu} h - g_{HZZ}^{(2)} Z_\nu \partial_\mu Z^{\mu\nu} h  \nonumber \\ & - & \frac{1}{2} g_{HWW}^{(1)} W^{\mu\nu} W^\dag_{\mu\nu} h - \Big[g_{HWW}^{(2)} W^\nu \partial^\mu W^\dag_{\mu\nu} h + {\rm h.c.} \Big]	\, ,
 \label{eq:anomL}
 \eeaa
with the relation between these anomalous coupling coefficients and the dimension-6 coefficients in our basis given by
\begin{align}
g^{(1)}_{hzz} &= \frac{2g}{c^2_W m_W} \left[ \bar{c}_{HB} s^2_W - 4\bar{c}_\gamma s^4_W + c^2_W \bar{c}_{HW} \right] \nonumber \\
g^{(2)}_{hzz} &= \frac{2g}{c^2_W m_W} \left[ (\bar{c}_{HW} + \bar{c}_W) c^2_W  + (\bar{c}_{HB} + \bar{c}_B)s^2_W \right] \nonumber \\
g^{(1)}_{hww} &=\frac{2g}{m_W}  \bar{c}_{HW}	 \nonumber	\\
g^{(2)}_{hww} &=\frac{g}{m_W}  (\bar{c}_W + \bar{c}_{HW})	\, .
\label{anomalousH}
\end{align}
We refer the reader to Ref.~\cite{benj} for more details and a complete list of Higgs anomalous couplings.

We calculate the effects of the dimension-6 operators on $V+H$ associated production by Monte-Carlo (MC) simulations
using {\tt MadGraph5~v2.1.0}~\cite{mg5} interfaced with {\tt Pythia}~\cite{pythia} and {\tt Delphes~v3}~\cite{delphes}, 
combined with the dimension-6 model implementation developed in~\cite{benj}. 
We start with $\bar{c}_W$ as an illustrative example, switching off all other coefficients, 
before considering briefly $\bar{c}_{HW}$ and then the full set of coefficients (\ref{eq:fullset}) in a global fit. 

\section{Kinematic Distributions in $H + V$ Production}

\subsection{The $H + V$ Invariant Mass Distribution Measured by D0}
\label{D0}

It was pointed out in~\cite{fastrack}, see also~\cite{ESY2}, that the invariant mass distribution in $H+V$ events
could be used to discriminate between minimally-coupled $J^P = 0^+, 0^-$ and 
graviton-like $2^+$ spin-parity assignments for the $H$ particle. Subsequently,
the D0 Collaboration has made available the observed $H + V$
invariant mass distribution as well as those expected in these scenarios~\cite{D0comb}. 
Here we use their background distribution and
simulate the signal events for a SM Higgs including the effects of non-zero dimension-6 coefficients, 
considering separately the 2-, 1- and 0-lepton channels for the decays  of vector bosons $V = Z, W^\pm$ 
produced in association with $H$ decaying to $b \bar{b}$.

\begin{figure}[h!]
\centering
\includegraphics[scale=0.8]{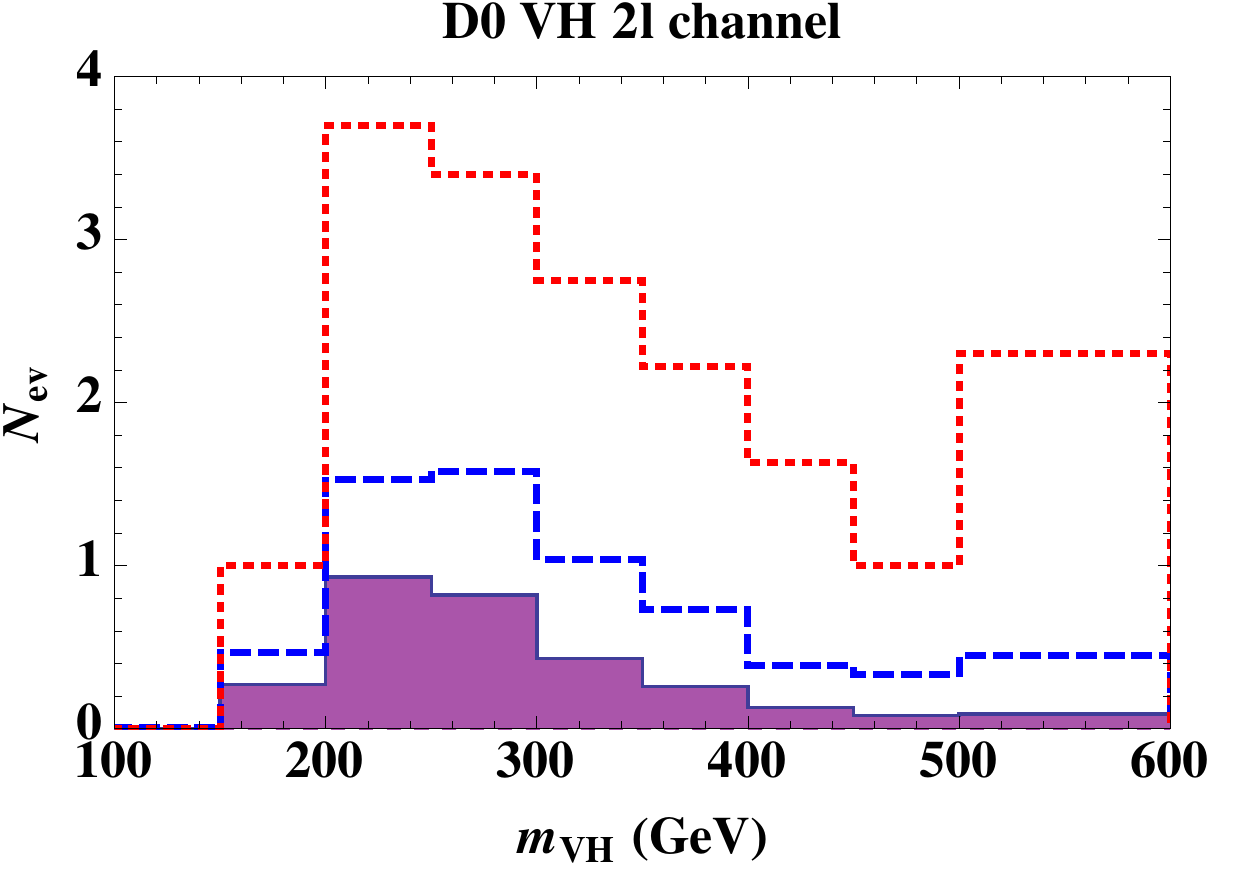}
\caption{\it Simulation of the $m_{VH}$ distribution in $(V \to 2 \ell) + (H \to {\bar b}b)$ events at the
Tevatron after implementing D0 cuts, obtained using {\tt MadGraph~v2.1.0} interfaced with {\tt Pythia} and {\tt Delphes~v3}, 
combined with the dimension-6 model implementation developed in~\protect\cite{benj}. 
The solid distribution is the SM expectation, while the red-dotted and blue-dashed lines 
correspond to the distributions with $\bar c_W$ =0.1 and 0.035, respectively.}
\label{mvh}
\end{figure}

Implementation details of the simulation can be found in Appendix~A. 
Summing the cross-section times efficiency over the 0-, 1- and 2-lepton channels,
we obtain the following signal strength as a function of $\bar{c}_W$ for $VH\to V \bar{b}b$ at D0,
\begin{equation*}
\mu_{H\bar{b}b} \simeq 1 + 29 \bar{c}_W	,
\end{equation*}
indicating a strong dependence of the signal strength on the coefficient of the
dimension-6 operator, which compensates for the relatively large error bar in the D0
measurement of this channel. We find that the best-fit signal strength $\mu_{H\bar{b}b} = 1.2  \pm 1.2$ 
reported by D0~\cite{D0comb} yields the following 95\% CL bounds in a $\chi^2$ fit:
\begin{equation*}
\bar{c}_W \in [-0.15 , 0.09 ] 	\, .
\end{equation*}
More information can be obtained from the differential kinematic distribution for $H + V$
production by considering the measurements in bins in $m_{VH}$,
which affords full sensitivity to $\bar{c}_W$ via the differential information 
available in the invariant mass distribution, particularly in the higher-mass bins where the 
signal-to-background ratio increases most rapidly. The invariant mass distribution found
in our simulation is plotted for the 2-lepton case in Fig.~\ref{mvh} for various values of $\bar{c}_W$. 
As expected, the effect of the dimension-6 operator is to generate a larger tail at high invariant masses than in the SM. 

\begin{figure}[h!]
\centering
\includegraphics[scale=0.35]{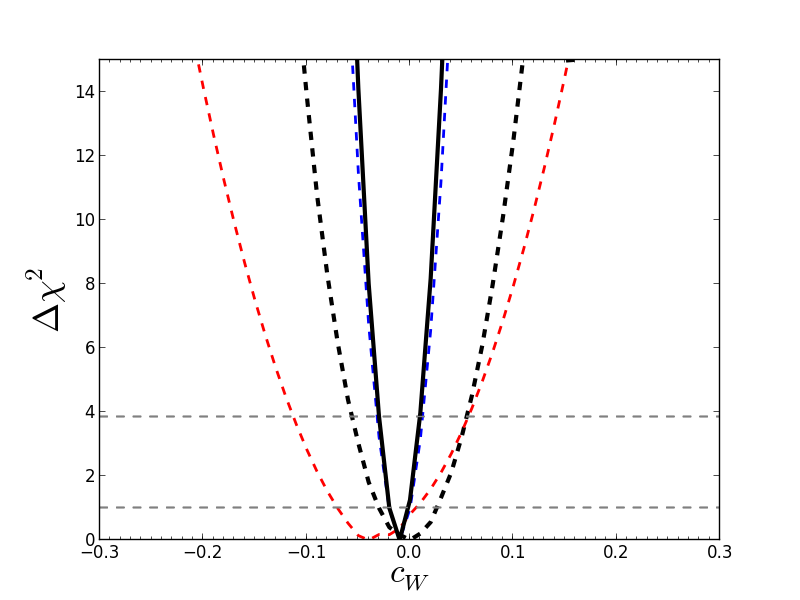}
\includegraphics[scale=0.35]{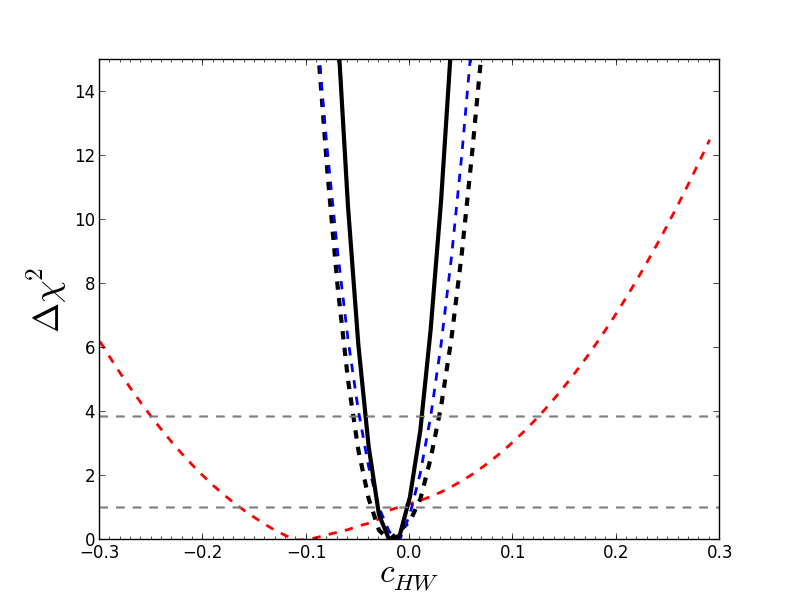}
\caption{\it The one-dimensional fit to the parameter $\bar c_W$ (left panel) and to $\bar c_{HW}$ (right panel). In each panel,
the dashed-red line corresponds to the constraint from the 0-, 1- and 2-lepton D0 $m_{VH}$ distribution including all bins, the 
dashed-blue line to the  0-, 1- and 2-lepton ATLAS $p_T^V$ distribution using the last bin only, the
dashed-black is the combination of CMS and ATLAS signal strengths in all channels except $VH$,
and the solid-black is the combination of all the above.}
\label{fig:cwandchw}
\end{figure}

We include the information from signal strength and differential distribution by constructing a $\chi^2$ function
with a contribution from each $m_{VH}$ bin.
We treat the errors provided as Gaussian, neglecting any correlations between bins
as this information is not available. Since the sensitivity of the distribution analysis is largely driven by the last bin, 
the sensitivity of the limit to correlations is minimal.  The resulting improved bounds are
\begin{equation}
\bar{c}_W \in [ -0.11 , 0.06 ] 	\, .
\label{D0mHV}
\end{equation}
The $\chi^2$ distribution from this constraint is shown as the dashed-red line in the left panel of Fig.~\ref{fig:cwandchw}.

This limit, using differential information, is better than the more inclusive observable $\mu_{HV}$ by  15-20 \%.
A better understanding of the tail in the kinematic distribution could improve considerably this limit. 
However, the Tevatron analysis is limited by statistics, whereas the LHC experiments benefit from increased energy,
which expands the available phase space and hence enhances the effect of anomalous couplings,
with the prospect also of future improvements in statistical significance. 
The study of constraints from Run~1 of the LHC at 8~TeV is the subject of the next section.

\subsection{The Vector-Boson Transverse-Momentum Distribution Measured by ATLAS}
\label{ATLAS}

\begin{figure}[h!]
\centering
\includegraphics[scale=0.35]{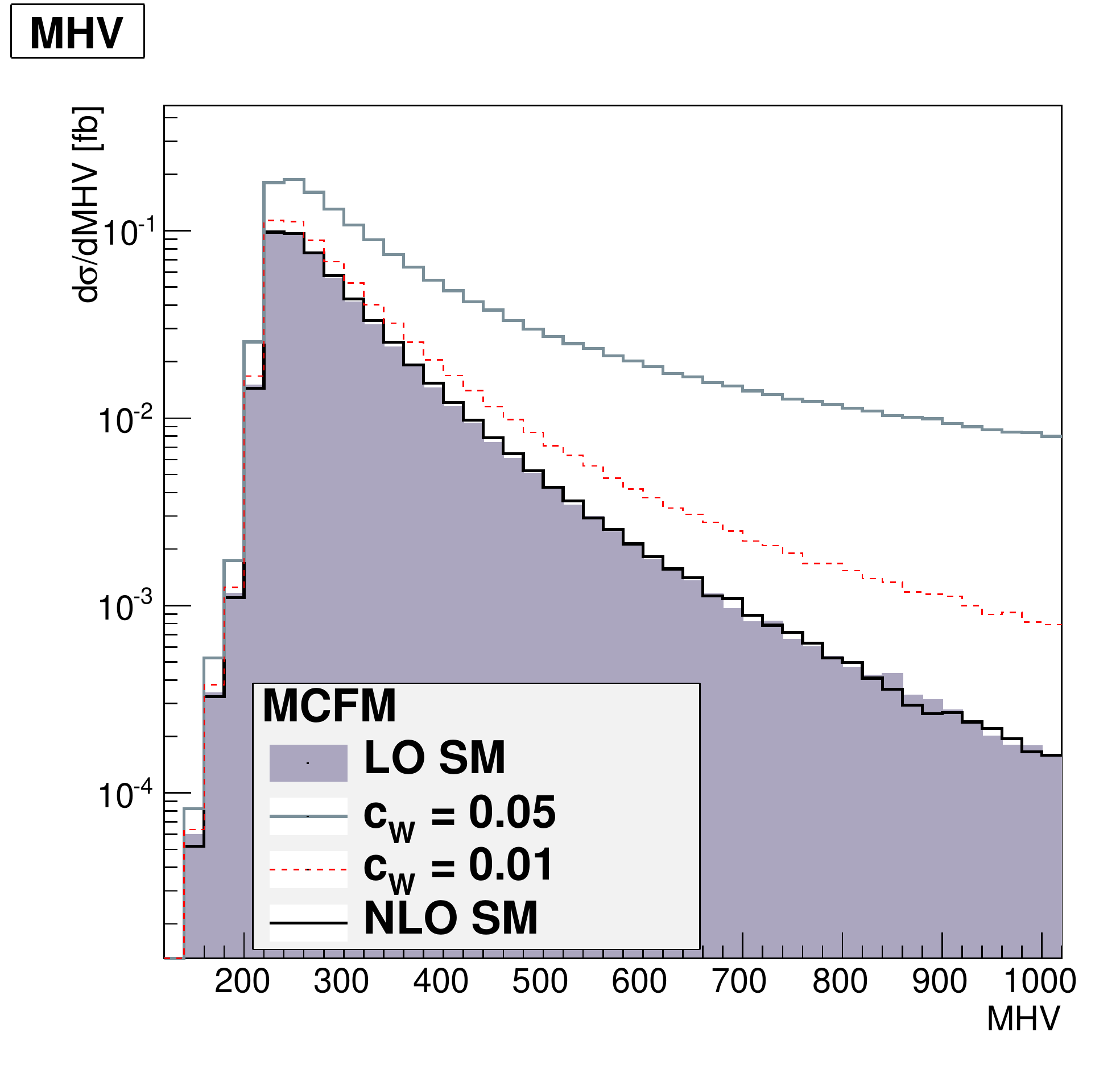}
\includegraphics[scale=0.35]{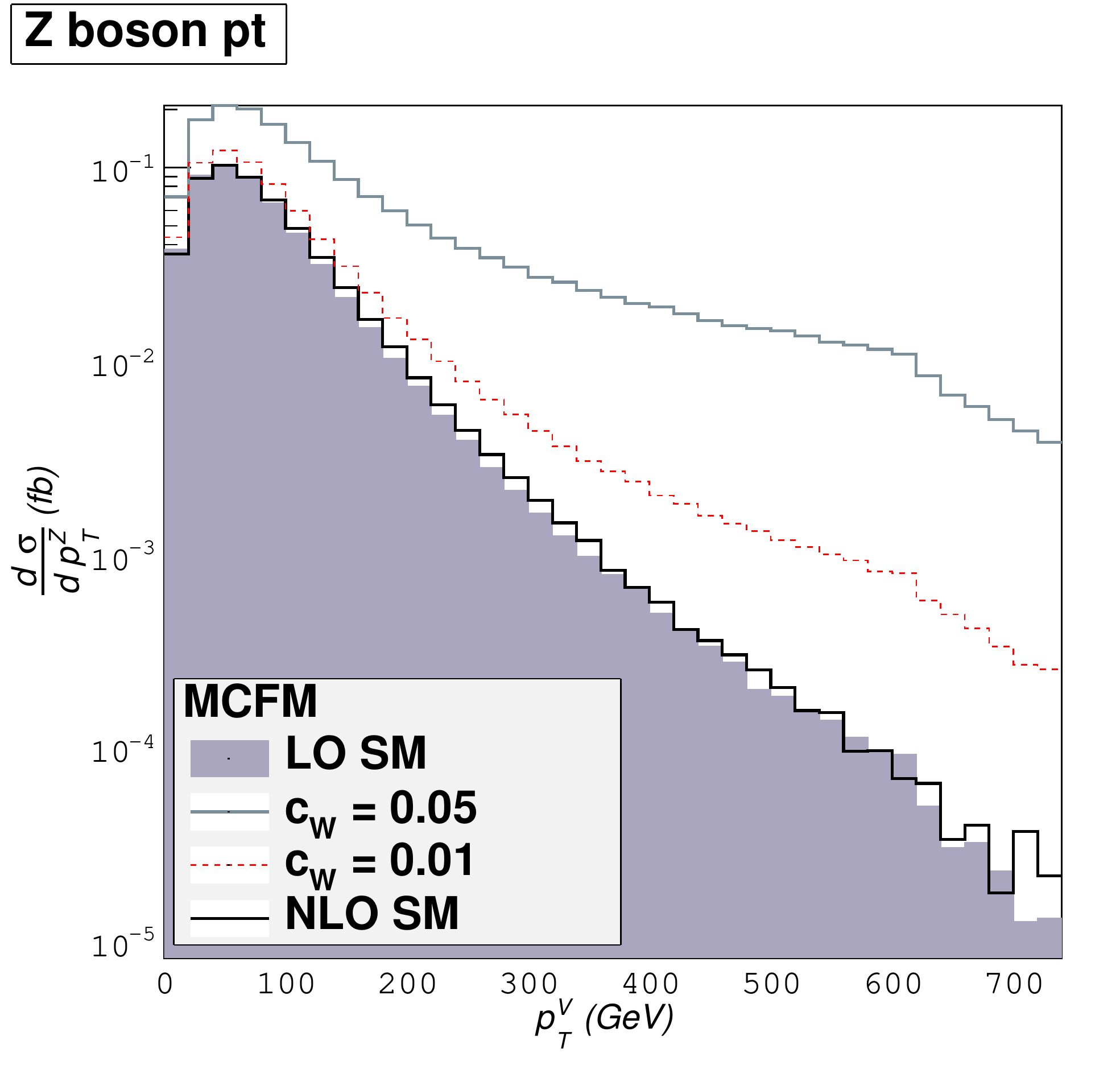}
\caption{\it The invariant mass (left panel) and transverse momentum (right panel)
distributions for LHC Run~1 at 8~TeV, calculated with LO and NLO QCD and compared with the effects of an effective operator.}
\label{NLO}
\end{figure}

The fact that dimension-6 operators generate a larger tail at higher invariant masses by 
modifying the production kinematics implies greater sensitivity at the LHC,
where the higher energy opens up the available phase space. Since the $V+H$ invariant mass distribution 
is not available, we make use here of the transverse momentum of the vector boson, $p_T^V$,
measured by ATLAS. However, the $p_T^V$ distribution is more affected by NLO 
QCD corrections than is the $V+H$ invariant mass distribution~\cite{ciaran}. 
We present in Fig.~\ref{NLO} the results of an NLO calculation using MCFM~\cite{MCFM}. 
Although the $p_T^V$ distribution is more sensitive to NLO corrections, the constraint on
the coefficient of an effective operator that we can obtain with LHC Run~1 data at 8~TeV
is still quite insensitive to the QCD higher order corrections. However, this will be an important
effect when reaching $\bar c_W \sim {\cal O}(10^{-3})$. Since such effects tend to broaden the 
$p_T^V$ distribution in the SM, the inclusion of NLO would only strengthen the bounds reported here
and as such will not modify our conclusions, which are reached under conservative assumptions. 

Details of the cuts implemented for the 0-,1- and 2-lepton ATLAS analysis can be found in Appendix~B.
Fig.~\ref{2lATLAS} is an example of the $p^T_V$ distribution for the 2-lepton signal in the bins used by the ATLAS search, 
for various values of $\bar{c}_W$.

\begin{figure}[h!]
\centering
\includegraphics[scale=0.8]{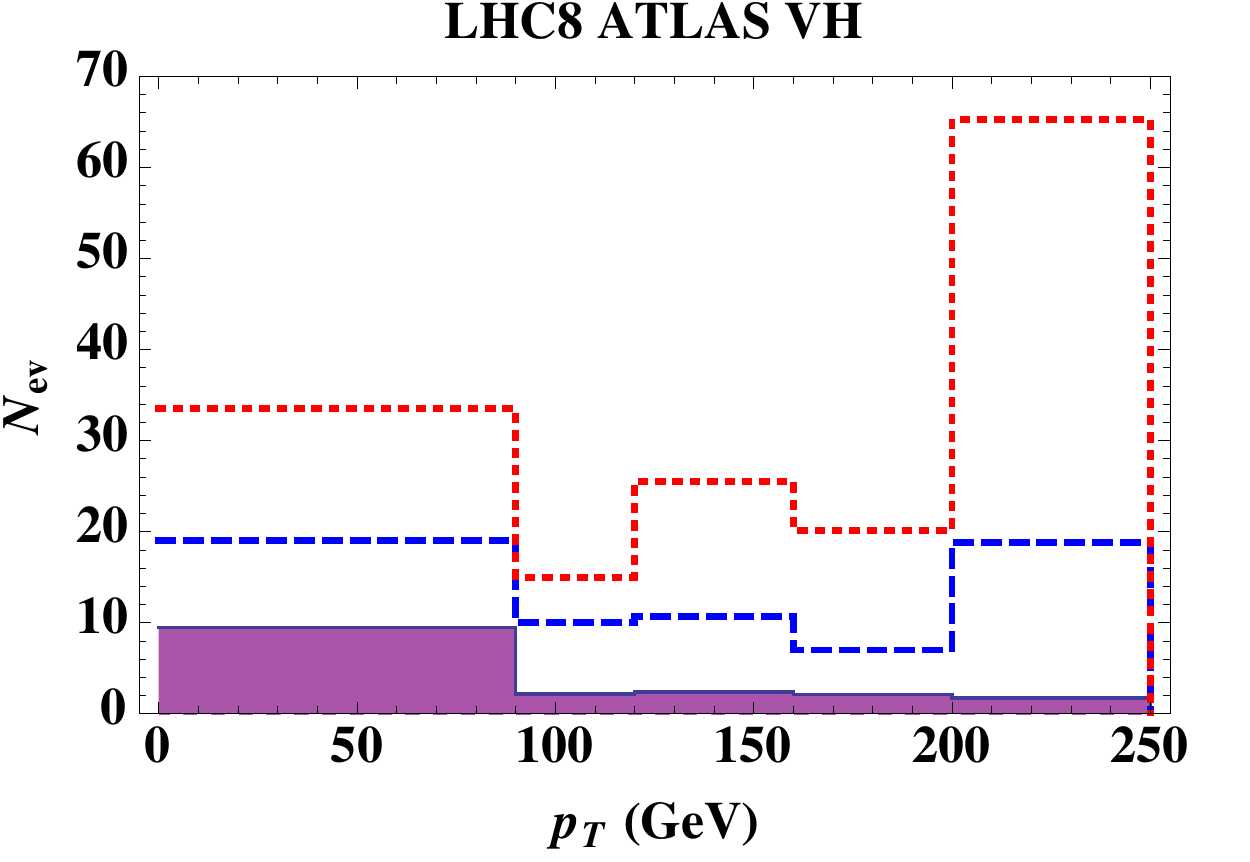}
\caption{\it Simulation of the $p_T^V$ distribution in $(V \to 2 \ell) + (H \to {\bar b}b)$ events at the
LHC after implementing ATLAS cuts, as obtained using {\tt MadGraph~v2.1.0} interfaced with {\tt Pythia} and {\tt Delphes~v3}, 
combined with the dimension-6 model implementation developed in~\protect\cite{benj}. 
The solid distribution is the SM expectation, and the red-dotted and blue-dashed lines 
correspond to the distributions with $\bar c_W$ =0.1 and 0.05, respectively.}
\label{2lATLAS}
\end{figure}

We see that the number of events in the last (overflow) bin increases rapidly with $\bar{c}_W$. 
Since the background overwhelms any signal in the lower bins, henceforth we
focus exclusively on this overflow bin where the signal-to-background ratio is highest.
A $\chi^2$ fit to the observed data gives the 95\% CL range
\begin{equation*}
\bar{c}_W \in [ -0.07, 0.07] 	\, ,
\end{equation*} 
which improves upon the D0 constraint (\ref{D0mHV}), as expected. The contribution to the $\chi^2$ function
from this constraint is shown as the dashed blue line in the left panel of Fig.~\ref{fig:cwandchw}.
For comparison, using the signal strength given for each of the 0-, 1- and 2-lepton channels,
which grow with $\bar{c}_W$ as
\begin{align*}
\mu_{2-lepton} &\simeq 1 + 23\bar{c}_W \\
\mu_{1-lepton} &\simeq 1 + 32\bar{c}_W \\
\mu_{0-lepton} &\simeq 1 + 33\bar{c}_W \, ,
\end{align*}
we find the 95\% CL range
\begin{equation*}
\bar{c}_W \in [ -0.09 , 0.03 ] \, ,
\end{equation*}
which is comparable to that using only the last bin of the $p^T_V$ differential distribution.

We emphasise that only the leading linear dependence on the dimension-6 coefficient is kept in our fit. 
Including the quadratic term could appear to give tighter constraints as it allows the signal to grow faster with increasing $\bar{c}_W$, 
but such bounds are spurious since it is not consistent to include a dependence on $\bar{c}_W^2$ without 
also introducing dimension-8 operators whose effects are formally of the same order. 
In the example given above, including the quadratic term would reduce the bounds to [ -0.06 , 0.03] 
for the signal-strength fit and $[ -0.04 , 0.04 ]$ for the binned fit. This sensitivity to higher-order effects 
indicates the level to which we may trust these constraints.
At the current level of precision,
the differences in the bounds between the linear and quadratic fits are larger than any uncertainties in
background distributions or MC simulations. 

Full results of one-dimensional fits for $\bar{c}_W$ are summarized on the left plot in Fig.~\ref{fig:cwandchw}.
In addition to the dashed red line corresponding to the analysis of the D0 $m_{VH}$ distribution
and the dashed blue line corresponding to the  ATLAS $p_T^V$ distribution discussed above, the
dashed black line is the combination of CMS and ATLAS signal strengths
including all channels except $VH$, and the solid black line is the combination of all the above. 
The right panel of Fig.~\ref{fig:cwandchw} shows the corresponding one-dimensional constraints on $\bar{c}_{HW}$,
where we see that the addition of the differential information is less important than for $\bar c_W$.

\section{Global Constraints From Signal Strengths and Differential Distributions}
\label{global}

Following these examples, we now combine the information from associated production measurements in the 
$H \to \bar{b}b$ final state by D0 and ATLAS together with the signal strengths in the 
$H \to \gamma\gamma, \gamma Z, WW, ZZ$ and $\tau\tau$ search channels measured by
CMS and ATLAS. We first constrain the dimension-6 coefficients individually, 
setting to zero all other coefficients, and then include the full set of coefficients (\ref{eq:fullset}) in a global fit.

The decay widths for $H\to Z^*Z^{(*)} \to 4l$, $H \to W^* W^{(*)} \to l\nu l\nu$, $H \to \bar{f}{f}$, 
$H\to gg$ and $H \to \gamma\gamma$ have dependences on the dimension-6 coefficients 
that are given in~\cite{silh-maggie}.
%
%
The dimension-6 operators also affect the vector boson fusion (VBF) production mode. 
Using the standard VBF cuts used at the LHC 8-TeV analysis,
namely $m_{jj} > $ 400 GeV, $p_T^j>  $ 20 GeV, $|\eta_j| <  $ 4.5 and $\Delta \eta_{jj}>$ 2.8, we find 
\begin{align*}
\frac{\sigma(p p \to V^* V^* j j \to h j j  )}{\sigma(p p \to V^* V^*  j j \to h  j j)_\text{SM}} &\simeq 1 - 8.30(\bar{c}_W + \text{tan}^2\theta_w \bar{c}_B) - 6.9(\bar{c}_{HW} + \text{tan}^2\theta_w \bar{c}_{HB}) -0.26\bar{c}_\gamma	\, .
\end{align*}
We confront these predictions with the likelihoods for the total signal strengths $\mu$ 
given by ATLAS and CMS in a particularly useful form~\cite{presentationLHCresults} 
as a 2-dimensional $\chi^2$ grid of $\mu_\text{ggF, tth}$ vs 
$\mu_\text{VBF,AP}$. For ATLAS we use the likelihoods made publicly available for diboson final states in~\cite{ATLASlikelihoods}
and the 2-dimensional $H \to \tau\tau$ likelihood given in~\cite{ATLAStautau}. 
The CMS likelihoods for the $H \to \gamma\gamma, WW^*, ZZ^*$ and $\tau\tau$ channels are taken from~\cite{CMSlikelihoods}. 
We assume gluon fusion and VBF to be the dominant production modes in all these channels, 
with associated production only entering the fit through the differential distributions of the D0 and ATLAS $\bar{b}b$ final states\footnote{The signal strength information is also included in the differential distribution through the normalisation of the heights of each bin to the total number of signal events.}.
The $H\to Z\gamma$ likelihood is reconstructed from the expected and observed 95\% CL signal strength 
using the method of~\cite{Contino}. 

\begin{figure}[h!]
\centering
\includegraphics[scale=0.35]{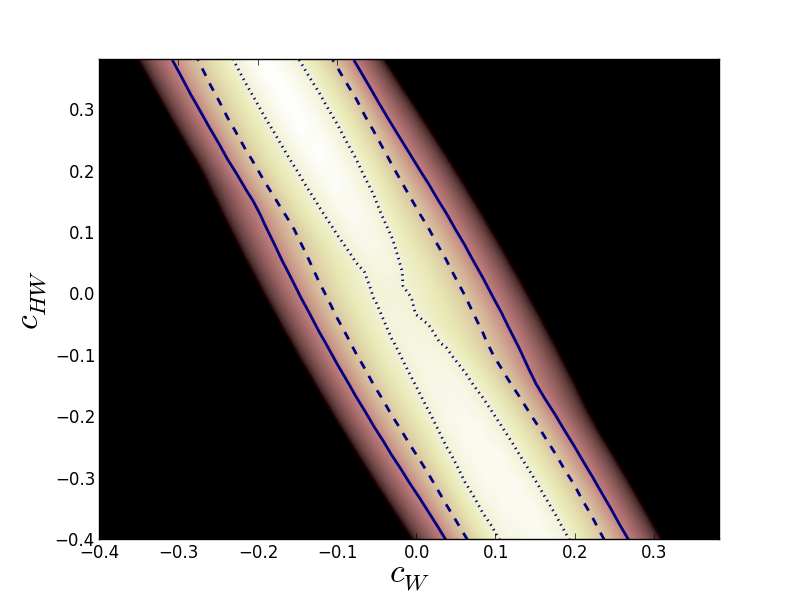}
\includegraphics[scale=0.35]{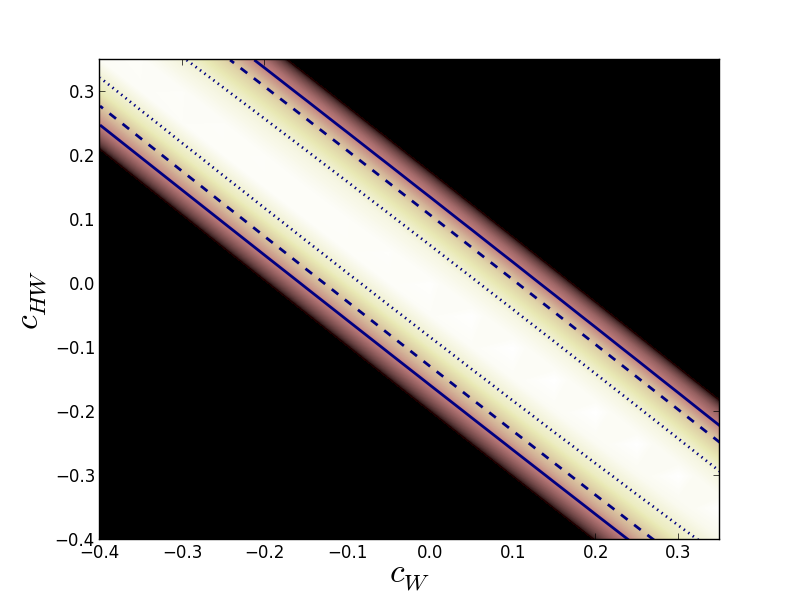}
\includegraphics[scale=0.35]{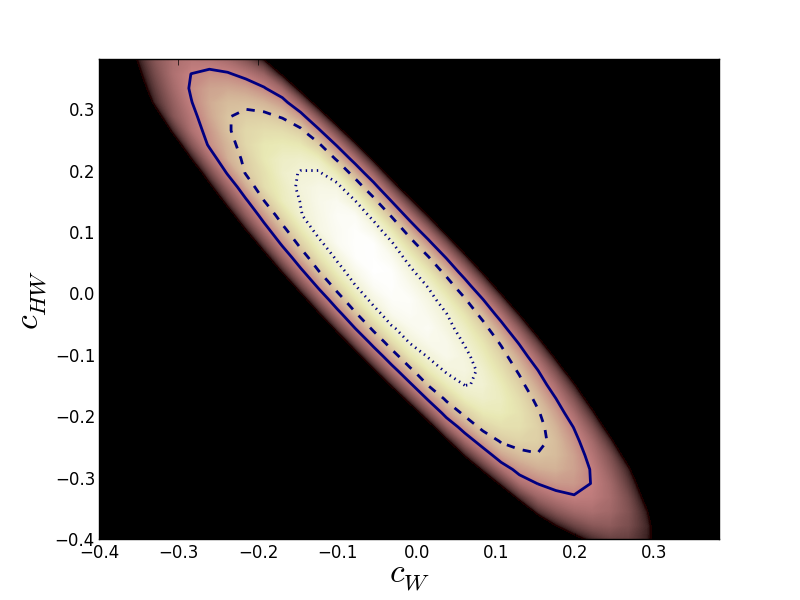}
\includegraphics[scale=0.35]{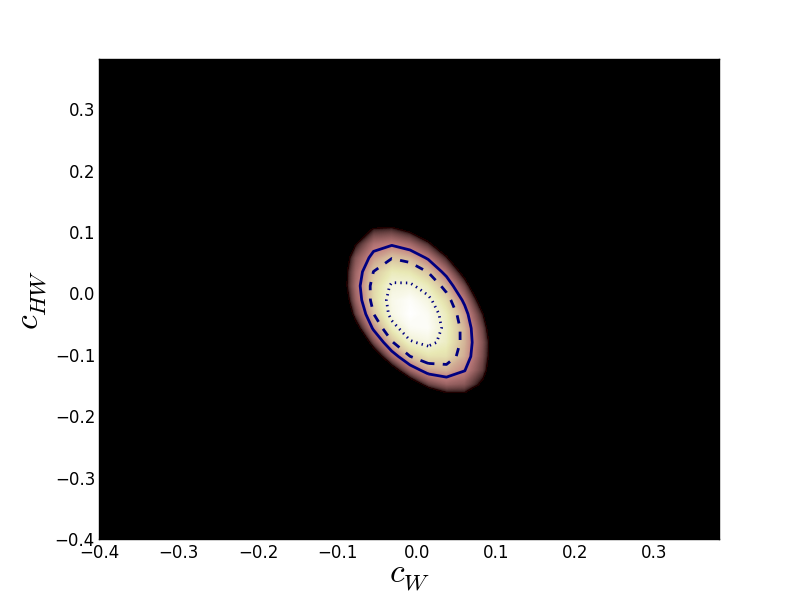}
\caption{\it Regions in the $({\bar c}_{W}, {\bar c}_{HW})$ planes allowed at the 68 (95) (99)\% CL
(in lighter shading and bounded by dotted, dashed and solid lines, respectively) in fits to the D0
$m_{VH}$ data alone (upper left panel), the ATLAS $p_T^V$ data alone (upper right panel), the
combination of these data (lower left panel) and a global fit using also signal-strength information
from CMS and ATLAS (lower right panel).}
\label{fig:cwvschw}
\end{figure}

\begin{figure}[h!]
\centering
\includegraphics[scale=0.35]{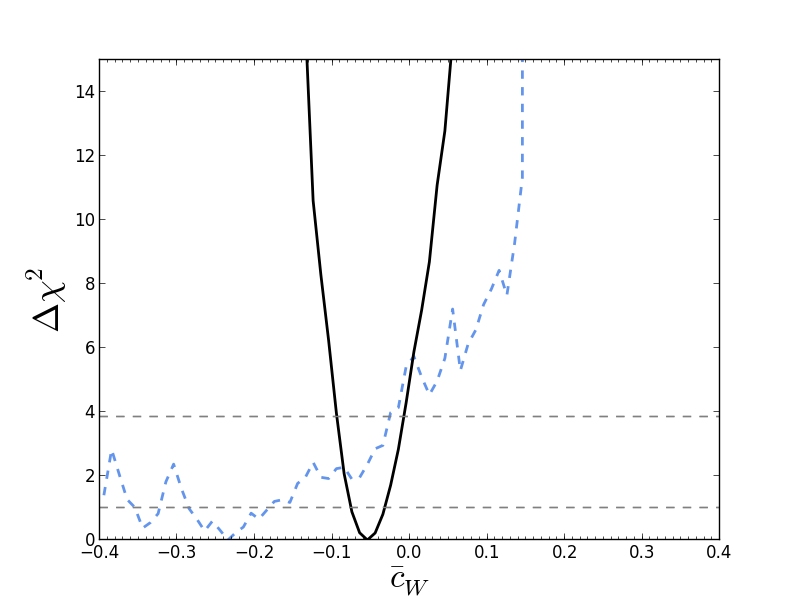}
\includegraphics[scale=0.35]{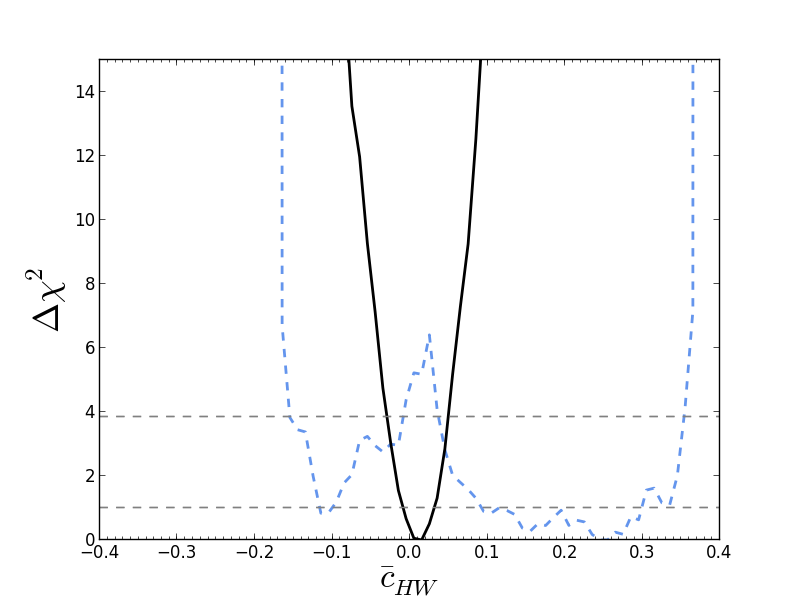}
\includegraphics[scale=0.35]{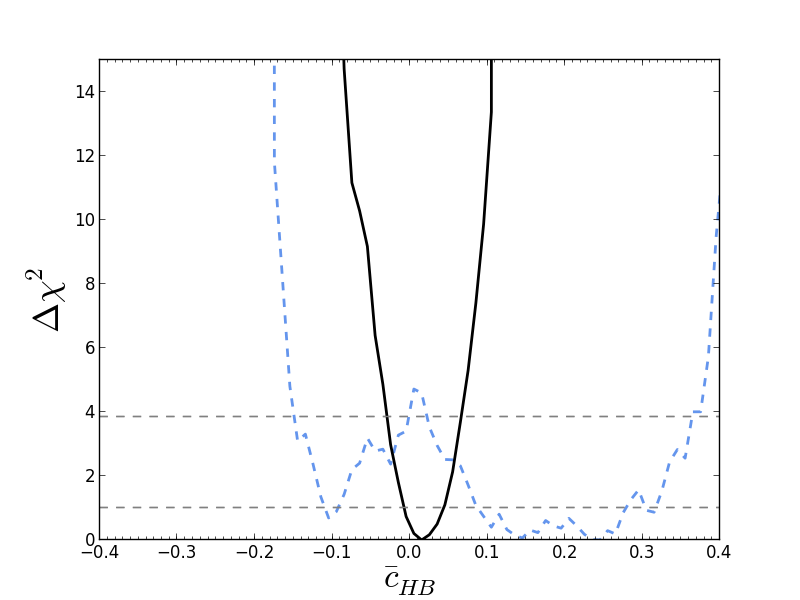}
\includegraphics[scale=0.35]{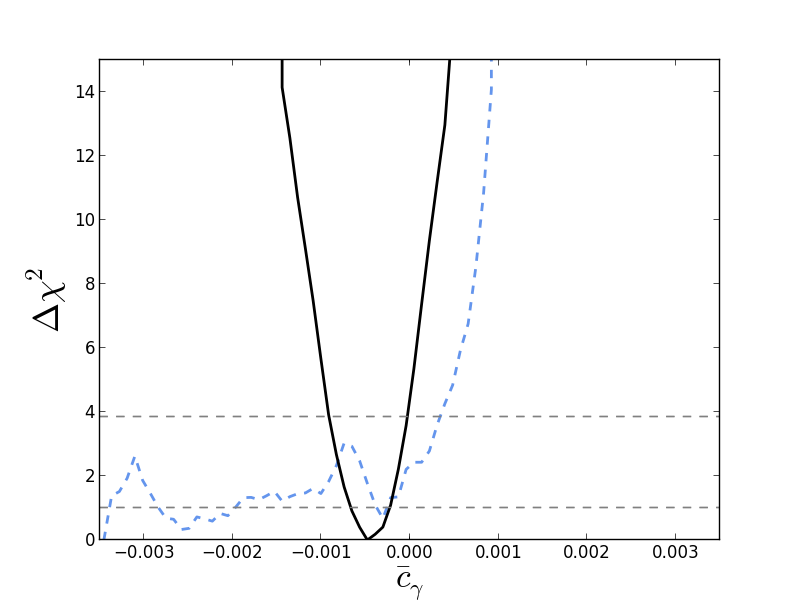}
\includegraphics[scale=0.35]{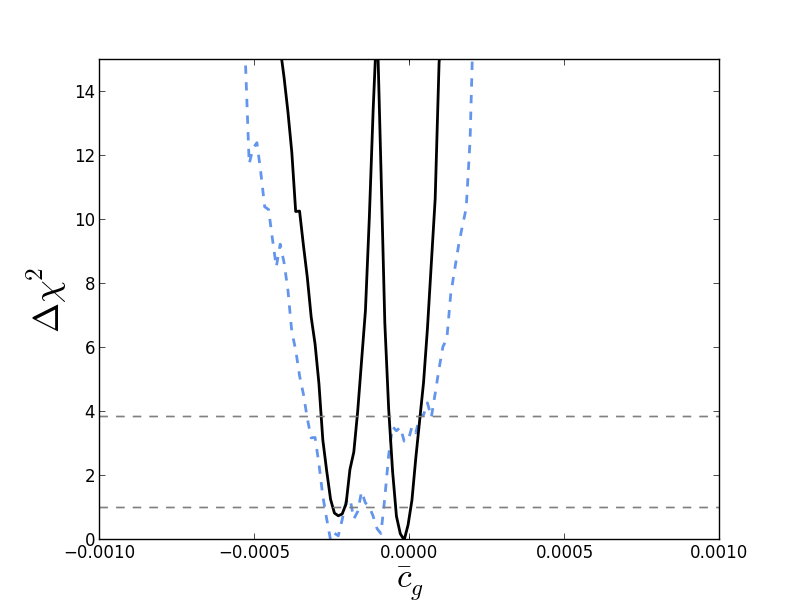}
\caption{\it Marginalized $\Delta\chi^2$ from a scan over the 8-dimensional parameter space (\protect\ref{eq:fullset}) using
the differential distribution information about $H+V$ associated production from D0 and ATLAS
as well as the ATLAS and CMS signal strengths (solid black line) and dropping the information from the kinematic
distributions (blue dashed line).}
\label{fig:scan}
\end{figure}

The result of the signal strength fit for all channels excluding $\bar{b}b$ at ATLAS and CMS gives the 
following 95\% CL range for $\bar{c}_W$, setting all other coefficients to zero:
\begin{equation*}
\bar{c}_W \in [-0.05 , 0.06]	\, .
\end{equation*}
Including the ATLAS $p^T_V$ and D0 $m_{VH}$ information discussed in the previous Section reduces this range to
\begin{equation*}
\bar{c}_W \in [-0.03 , 0.01]	\, .
\end{equation*}
The improvement of the limit on a single operator is significant. Furthermore the importance of using as many inputs as possible becomes clear when one includes several operators
simultaneously~\cite{eduard}. For example, allowing the coefficient ${\bar c}_{HW}$ to vary simultaneously
with $\bar{c}_W$ introduces a possible
degeneracy in the fit, as shown in the upper left panel of Fig.~\ref{fig:cwvschw}. We see that the D0 $m_{VH}$ data
alone constrain essentially just one linear combination of ${\bar c}_{W}$ and ${\bar c}_{HW}$, and a similar
effect occurs in the upper right panel where the result of a 2-parameter fit to just the ATLAS $p_T^V$ data
is shown. However, the correlation coefficients are somewhat different, so that combining the two sets of data breaks the degeneracy
to some extent, as seen in the lower left panel of Fig.~\ref{fig:cwvschw}. Finally, in the lower right panel of
Fig.~\ref{fig:cwvschw} the degeneracy between $\bar{c}_W$ and $\bar{c}_{HW}$ is completely removed
when the D0 and ATLAS associated production data are combined
with the signal strength data from the other channels. This is primarily because, of the two operators
considered here, only $\bar{c}_{W}$ enters in the $H\to\gamma\gamma$ decay width. 

\begin{figure}[h!]
\centering
\includegraphics[scale=0.7]{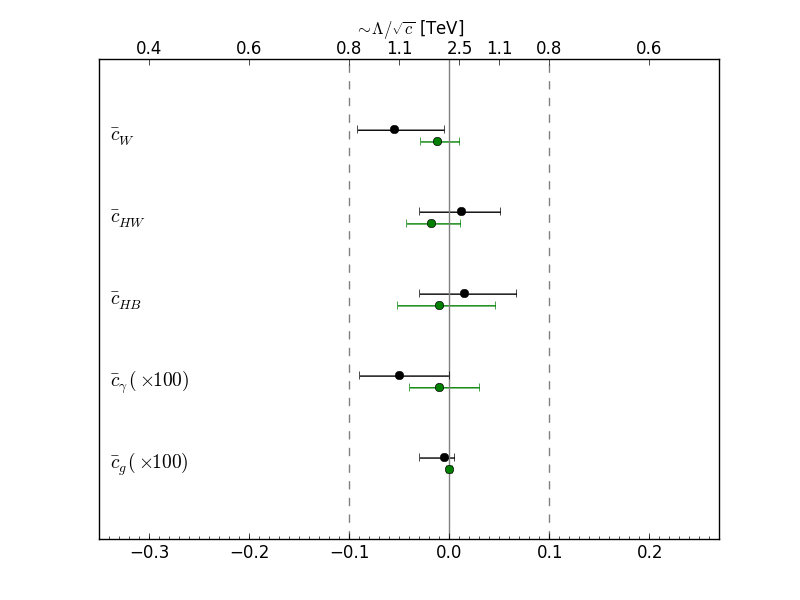}
\caption{\it The 95\% CL ranges allowed in a global fit to the dimension-6 operator coefficients listed in (\ref{eq:fullset}) (black),
and the 95\% CL ranges allowed for each operator coefficient individually, setting the others to zero (green). The upper axis is the corresponding sensitivity to the scale $\Lambda/\sqrt{c}$ in TeV where $\bar{c} \equiv c\frac{v^2}{\Lambda^2}$. Note that $\bar{c}_{\gamma,g}$ are shown $\times 100$ for which the upper axis should therefore be read $\times 10$.} 
\label{fig:summary}
\end{figure}

Finally we consider the full set of 8 dimension-6 operators listed in (\ref{eq:fullset}), 
setting $c_b = c_\tau \equiv c_d$, including a linear dependence on these coefficients in the ATLAS and CMS signal strengths,
combined with the differential distribution information of $H+V$ associated production at ATLAS and D0
discussed in Section~3. The result of a scan over the 8-dimensional parameter space is represented by the marginalized $\Delta\chi^2$ in solid black in 
Fig.~\ref{fig:scan}. The blue dashed line in Fig.~\ref{fig:scan} is the result of the 8-parameter fit using only 
ATLAS and CMS signal strengths without $H + V \to V \bar{b} b$ associated production information.
We see that omitting associated production yields no significant constraints on any of the operators aside from $\bar{c}_g$\footnote{The 
bi-modal distribution of $\bar{c}_g$ is due to the linear dependence on the coefficient of the gluon production 
cross-section rescaling, which is not allowed to go negative and so is responsible for the two minima in the best fit.}.

The scan over the 8-dimensional parameter space including the kinematical information from $H + V$ production yields
the 95\% CL bounds summarized in the black error bars of Fig.~\ref{fig:summary}.
Also shown in green in Fig.~\ref{fig:cwandchw} are the 1-dimensional constraints obtained by switching on one operator at a time
with all others set to zero. We omit $c_t, c_d$ and $c_H$ in this and the previous figure,
as no meaningful constraints are found for these coefficients. 

\begin{figure}[h!]
\centering
\includegraphics[scale=0.7]{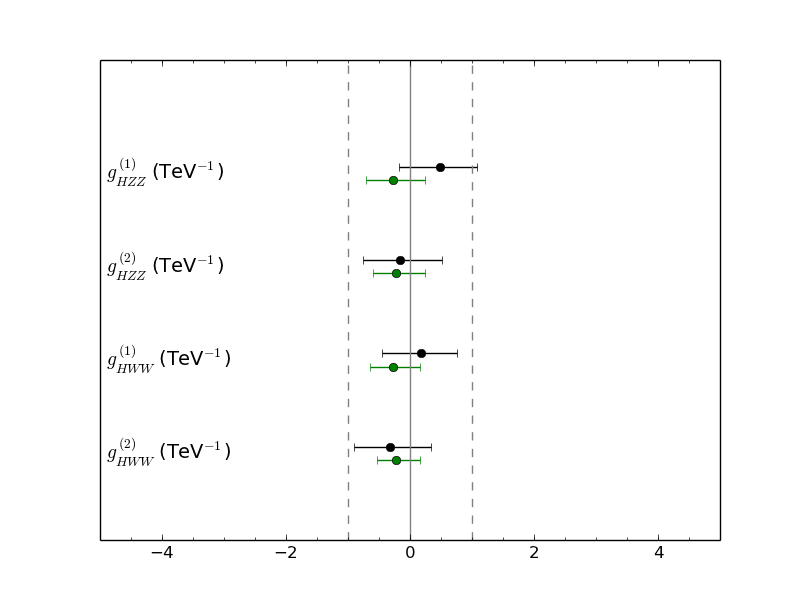}
\caption{\it The 95\% CL ranges allowed in a global fit to the anomalous Higgs couplings listed in (\ref{anomalousH}) (black),
and the 95\% CL ranges allowed for each coupling individually, setting the others to zero (green).} 
\label{fig:summaryg}
\end{figure}

We may also express the bounds obtained here in terms of the Higgs anomalous couplings
as parametrized in (\ref{eq:anomL}). Our results are displayed in Fig.~\ref{fig:summaryg}
using the same colour coding as in Fig.~\ref{fig:summary}.


\section{Conclusions}\label{conx}

With Higgs property measurements consistent with SM expectations,
and no clear sign of new physics from Run I of the LHC, it is natural to consider the SM as an 
effective theory supplemented by dimension-6 operators whose effects are suppressed by the scale of new physics. 
In this model-independent approach it is particularly interesting to consider a complete set of operators that 
minimizes any assumptions on the Wilson coefficients one chooses to include, 
thus providing truly universal bounds if one accepts the framework of the SM and decoupled new physics. 

In this analysis we considered the set of CP-even operators that affect the Higgs sector at tree-level. 
Certain operators contain derivative interactions that modify the kinematics in $H+V$ associated production, 
modifying in particular the tail in the differential distribution of the $V+H$ invariant mass and the 
vector boson transverse momentum. We simulated the $V+H \to Vb\bar{b}$ process at D0 and
found greater sensitivity to dimension-6 operators using the differential invariant mass distribution
than using only signal strength information in this channel. Since the higher energies of the LHC
enlarge the available phase space for boosted new physics, observations of the same process
by ATLAS and CMS are expected to be more sensitive than D0 to the effects of dimension-6 operators,
as we have confirmed here. Moreover, including kinematic distributions from both Tevatron and LHC can help remove degeneracies in multi-parameter fits. 

Including differential distributions of associated production with the signal strength from other channels, 
we have performed a scan of the 8-dimensional parameter space of the CP-even dimension-6 operator coefficients and 
placed 95\% CL bounds. Without the use of associated production information, there are
degeneracies that give flat directions in the fit. These could otherwise be eliminated using measurements of TGCs. 
However, this may introduce model-dependent assumptions as TGCs, despite their
greater sensitivity compared to Higgs measurements, also contain a poorly constrained direction due to a partial cancellation among contributions to $e^+e^- \to W^+ W^-$. 
Thus the use of associated Higgs production complements other ingredients
in global fits to a complete set of operators. As better measurements of TGCs at the LHC become available it will be interesting to fully explore this complementarity, which we intend to address in future work. This information will grow in importance when
higher-energy LHC data are analyzed, since the increased phase space will further improve the sensitivity
to dimension-6 operators.

\emph{Note added:} We thank A. Knochel and the authors of Ref.~\cite{BKKLR} for pointing out to us that the previous version of this paper underestimated the ATLAS $p^T_V$ constraints due to a misinterpretation of the expected number of SM events in Table 5 of Ref.~\cite{atlasptv}, which actually corresponds to a best fit signal strength of 0.2. Normalising instead to a signal strength of 1.0 yields improved constraints competitive with those of LEP, in agreement with comparable results in \cite{BKKLR}.

 \section*{Acknowledgements}

VS thanks Adam Falkowski, Alex Pomarol, Francesco Riva and Ciaran Williams and TY thanks Robert Hogan and Thomas Richardson for useful conversations. The authors are grateful to Jonathan Hays for providing information on the D0 analysis.
The work of JE was supported partly by the London
Centre for Terauniverse Studies (LCTS), using funding from the European
Research Council via the Advanced Investigator Grant 267352. The
work of VS is supported by the STFC grant ST/J000477/1.
The work of TY was supported by a Graduate Teaching Assistantship from
King's College London.

 \appendix
 \label{appD0}
 
 \section{D0 $H+V$ Analysis}
 
 \subsection{$p \bar{p} \to Z h \to l \bar{l} b \bar{b}$}
 
The event selection for the 2-lepton channel is taken from~\cite{D02lepton}. 
The basic cuts for di-electrons are $p_T > 15$, $|\eta| < 15$ and at least one electron with $|\eta| < 1.1$, 
and for di-muons are $p_T > 10$ GeV, $|\eta| < 2$ and at least one muon with $p_T > 15 \text{GeV}, |\eta|<1.5$. 
The muons have an isolation cut that requires them to be separated from all jets by 
$\Delta R = \sqrt{\Delta\eta^2 + \Delta\phi^2} > 0.5$. 

The ``pretag'' cuts are then applied to keep only events with $70 < M_{ll} < 110$ GeV 
and at least two jets having $p_T > 20$ GeV and $|\eta| < 2.5$. The final selection step is $b$-tagging the jets 
according to ``loose'' and ``tight'' categories, with at least one tight and one loose $b$-tagged jet. 
We simulate this double-tagged (DT) requirement by using the efficiencies reported as a function of
$p_T$ in~\cite{D0jet}. Fitting to Fig.~6a and 6b in that reference yields the following 
formula for the loose and tight efficiencies $\epsilon$:
\begin{align*}
\epsilon_\text{loose} &= a_\text{loose}e^{-\frac{p_T}{600}} \text{tanh}(0.020p_T + 0.77) \, ,\\
\epsilon_\text{tight} &= a_\text{tight}e^{-\frac{p_T}{360}} \text{tanh}(0.029p_T + 0.34) \, ,
\end{align*}
where the coefficients $a_\text{loose} = 0.79, a_\text{tight}=0.70$ in the region $|\eta| < 1.5$ 
and $a_\text{loose} = 0.67, a_\text{tight}=0.58$ for $|\eta| > 1.5$, 
the efficiency being fairly flat as a function of $\eta$ in these regions. 

Finally we set the {\tt Delphes} ECAL and HCAL resolutions as functions of energy 
$E$ to $0.01E + 0.2\sqrt{E} + 0.25$ and $0.050E + 0.8\sqrt{E}$ respectively. 
The same expression is used for the ECAL electron energy resolution.

After running our simulation we obtain the number of signal events by multiplying the cross-section given by 
{\tt MadGraph} with the efficiency after cuts and reweighting by a k-factor of 1.5 as an overall normalization. 
We find the resulting number of pretag and DT signal events for a SM Higgs to be 8.6 and 3.1 respectively, 
in agreement with the numbers listed in Table 3 of~\cite{D02lepton}. We have
also verified that we reproduce well the distribution of $H + V$ invariant masses for the SM Higgs signal
given by D0 in Fig.~2c of~\cite{D0spin}. 

\subsection{$p \bar{p} \to W h \to l \nu b \bar{b}$} 

We implement the cuts listed in~\cite{D01lepton} by requiring one electron (muon) with $p_T > 15$ and $|\eta| < 2.5$ $(2.0)$, 
and by requiring two jets with $p_T > 20$ GeV and $|\eta| < 2.5$. The muon is required to be isolated from all jets by 
$\Delta R > 0.5$. Finally, the transverse mass $M^W_T$, defined as  
$2p^l_T \slashed{E}_T (1 - \text{cos}\Delta\phi(l, \slashed{E}_T))$, must satisfy $M^W_T > 40 \text{GeV} - 0.5\slashed{E}_T$. 
This defines the pretag events with the $b$-tag cut then applied as described previously. 
Running the simulation with the cross-section times efficiency reweighted by a k-factor of 1.7 gives
good agreement with the expected numbers of pretag and final events given in Table~1 of~\cite{D01lepton}.

\subsection{$p \bar{p} \to Z h \to \nu \bar{\nu} b \bar{b}$}

Following~\cite{D0zerolepton}, we select events containing two jets with $p_T > 20$~GeV and $|\eta| < 2.5$, 
whose opening angle is $\Delta\phi < 165^\circ$, and apply a missing transverse energy cut $\slashed{E}_T > 40$ GeV. 
The jets are furthermore required to have the scalar sum of the their transverse momenta larger than 80 GeV. 
We also reject events with an isolated muon or electron having $p_T > 15$~GeV. 
We verified that the resulting numbers of events both before and after $b$-tag cuts
agree within errors with the numbers given in Table~1 of~\cite{D0zerolepton} without any reweighting. 

\section{ATLAS $H+V$ Analysis}

The implementation of this analysis follows the cuts given in \cite{atlasptv}. 
 
\subsection{$p \bar{p} \to Z h \to l \bar{l} b \bar{b}$}

We select events with exactly 2 muons (electrons) satisfying $|\eta| < 2.5 \, (2.47)$ and $83 < M_{ll} < 99$ GeV. A missing transverse energy cut of $E^\text{miss}_T$ is applied. There must be only 2 b-tagged jets with the higher-$p_T$ jet $ > 45$ GeV and $p_T  > 20$ GeV for the other jet, and both with $|\eta| < 2.5$. Finally we place a $\Delta R$ cut on the angle between the two jets which varies depending on the $p^V_T$ bin (see Table 2 in \cite{atlasptv}). The transverse momentum $p^V_T$ of the vector boson is reconstructed using the vector sum of the transverse components of the two leptons. 

We simulate events at the 8 TeV LHC with the resulting distribution in the $p^V_T$ bins  used by ATLAS. 
We reweight the cross-section so as to normalise the number of signal events in each bin to the expected SM count from Table 5 of \cite{atlasptv}.

\subsection{$p \bar{p} \to W h \to l \nu b \bar{b}$} 

In this sub-channel we select exactly one muon (electron) with $|\eta| < 2.5 (2.47)$ and $E_T > 25$ GeV. The missing transverse energy requirement is $E^\text{miss}_T > 25 \, (50)$ for $p^V_T$ less (greater) than 200 GeV. The invariant transverse mass $m^W_T$ is required to be less than 120 GeV, and for $p^V_T < 160$ GeV it must also be greater than 40 GeV. The $p^V_T$ transverse momentum is in this case the vector sum of the transverse components of the lepton and missing $E_T$. The jet requirements are the same as for the 2-lepton case, and we have normalised our number of events after simulation in the same way as above.

\subsection{$p \bar{p} \to Z h \to \nu \bar{\nu} b \bar{b}$}

Here we require no leptons that pass the other criterias and a large missing transverse energy of $E^\text{miss}_T > 120$ GeV with $p^\text{miss}_T > 30$ GeV and an angle between the two of $\Delta\phi < \pi/2$. The azimuthal angle between the $E^\text{miss}_T$ and the vector sum of the jets must be $\Delta\phi > 4.8$, as well as $\Delta\phi > 1.5$ with the nearest jet. The other jet cuts and  $\Delta R$ requirements as a function of $p^V_T$ are also the same here, with the $p^V_T$ identified as the $E^T_\text{miss}$.


\begin{thebibliography}{10}


\bibitem{Eureka}
G.~Aad {\it et al.}  [ATLAS Collaboration],
  Phys.\ Lett.\ B {\bf 716} (2012) 1
  [arXiv:1207.7214 [hep-ex]];
S.~Chatrchyan {\it et al.}  [CMS Collaboration],
  Phys.\ Lett.\ B {\bf 716} (2012) 30
  [arXiv:1207.7235 [hep-ex]].
 
\bibitem{strengths}
S.~Chatrchyan {\it et al.}  [CMS Collaboration],
  JHEP {\bf 1306} (2013) 081
  [arXiv:1303.4571 [hep-ex]].
  ATLAS Collaboration, ATLAS-CONF-2014-010, {\tt http://cds.cern.ch/record/1670531/files/ATLAS-CONF-2014-010.pdf}.
  
\bibitem{fits}
 M.~Baak, M.~Goebel, J.~Haller, A.~Hoecker, D.~Ludwig, K.~Moenig, M.~Schott and J.~Stelzer,
  Eur.\ Phys.\ J.\ C {\bf 72} (2012) 2003
  [arXiv:1107.0975 [hep-ph]].
D.~Carmi, A.~Falkowski, E.~Kuflik and T.~Volanski,
arXiv:1202.3144 [hep-ph];
A.~Azatov, R.~Contino and J.~Galloway,
  JHEP {\bf 1204} (2012) 127  [hep-ph/1202.3415].
J.R.~Espinosa, C.~Grojean, M.~Muhlleitner and M.~Trott,
arXiv:1202.3697 [hep-ph];
P.~P.~Giardino, K.~Kannike, M.~Raidal and A.~Strumia,
arXiv:1203.4254 [hep-ph];
T.~Li, X.~Wan, Y.~Wang and S.~Zhu,
arXiv:1203.5083 [hep-ph];
M.~Rauch,
arXiv:1203.6826 [hep-ph];
 J.~Ellis and T.~You,
JHEP {\bf 1206} (2012) 140,
[arXiv:1204.0464 [hep-ph]];
A.~Azatov, R.~Contino, D.~Del Re, J.~Galloway, M.~Grassi and S.~Rahatlou,
arXiv:1204.4817 [hep-ph];
M.~Klute, R.~Lafaye, T.~Plehn, M.~Rauch and D.~Zerwas,
arXiv:1205.2699 [hep-ph];
L.~Wang and X.-F.~Han, 
Phys. Rev. D {\bf 86} (2012) 095007, 
[arXiv:1206.1673 [hep-ph]]; 
 D.~Carmi, A.~Falkowski, E.~Kuflik and T.~Volansky,
 arXiv:1206.4201 [hep-ph]; 
 M.~J.~Dolan, C.~Englert and M.~Spannowsky,
 arXiv:1206.5001 [hep-ph];
 J.~Chang, K.~Cheung, P.~Tseng and T.~Yuan,
 arXiv:1206.5853 [hep-ph];
 S.~Chang, C.~A.~Newby, N.~Raj and C.~Wanotayaroj,
 arXiv:1207.0493 [hep-ph];
 I.~Low, J.~Lykken and G.~Shaughnessy,
 arXiv:1207.1093 [hep-ph];
  J.~Ellis and T.~You,
  JHEP {\bf 1209}, 123 (2012)
  [arXiv:1207.1693 [hep-ph]].
P.~P.~Giardino, K.~Kannike, M.~Raidal and A.~Strumia, 
 arXiv:1207.1347 [hep-ph];
M.~Montull and F.~Riva,
arXiv:1207.1716 [hep-ph];
J.~R.~Espinosa, C.~Grojean, M.~Muhlleitner and M.~Trott,
arXiv:1207.1717 [hep-ph];
D.~Carmi, A.~Falkowski, E.~Kuflik, T.~Volansky and J.~Zupan, 
arXiv:1207.1718 [hep-ph];
S.~Banerjee, S.~Mukhopadhyay and B.~Mukhopadhyaya,
JHEP {\bf 10} (2012) 062,
[arXiv:1207.3588 [hep-ph]];
F.~Bonner, T.~Ota, M.~Rauch and W.~Winter,
arXiv:1207.4599 [hep-ph];
T.~Plehn and M.~Rauch,
arXiv:1207.6108 [hep-ph];
A.~Djouadi,
arXiv:1208.3436 [hep-ph];
B.~Batell, S.~Gori and L.~T.~Wang,
arXiv:1209.6832 [hep-ph];
G.~Moreau,
Phys.~Rev.~D {\bf 87} (2013) 015027,
[arXiv:1210.3977 [hep-ph]];
G.~Cacciapaglia, A.~Deandrea, G.~D.~La Rochelle and J-B.~Flament,
arXiv:1210.8120 [hep-ph];
E.~Masso and V.~Sanz,
arXiv:1211.1320 [hep-ph];
R.~Tito D'Agnolo, E.~Kuflik and M.~Zanetti,
arXiv:1212.1165 [hep-ph];
A.~Azatov and J.~Galloway,
arXiv:1212.1380 [hep-ph];
G.~Bhattacharyya, D.~Das and P.B.~Pal,  
Phys.~Rev.~D {\bf 87} (2013) 011702,
[arXiv:1212.4651 [hep-ph]];
D.~Choudhury, R.~Islam, A.~Kundu and B.~Mukhopadhyaya,
arXiv:1212.4652 [hep-ph];
R.~S.~Gupta, M.~Montull and F.~Riva,
arXiv:1212.5240 [hep-ph];
G.~Belanger, B.~Dumont, U.~Ellwanger, J.~F.~Gunion and S.~Kraml,
arxiv:1212.5244 [hep-ph];
K.~Cheung, J.~S.~Lee and P-Y.~Tseng,
arXiv:1302.3794 [hep-ph].
A.~Falkowski, F.~Riva and A.~Urbano,
arXiv:1303.1812 [hep-ph],
P.~P.~Giardino, K.~Kannike, I.~Masina, M.~Raidal and A.~Strumia,
arXiv:1303.3570 [hep-ph].
 J.~Ellis and T.~You,
  JHEP {\bf 1306} (2013) 103
  [arXiv:1303.3879 [hep-ph]].
 S.~Banerjee, S.~Mukhopadhyay and B.~Mukhopadhyaya,
  Phys.\ Rev.\ D {\bf 89}, 053010 (2014)
  [arXiv:1308.4860 [hep-ph]].
 
  
\bibitem{primafacie}
  J.~Ellis, V.~Sanz and T.~You,
  Phys.\ Lett.\ B {\bf 726} (2013) 244
  [arXiv:1211.3068 [hep-ph]].
  
 \bibitem{spinparity}
 J.~Ellis and D.~S.~Hwang,
  JHEP {\bf 1209} (2012) 071
[arXiv:1202.6660 [hep-ph]];
  A.~Alves,
  arXiv:1209.1037 [hep-ph];
J.~Ellis, R.~Fok, D.~S.~Hwang, V.~Sanz and T.~You,
arXiv:1210.5229 [hep-ph];
see also Appendix~A of
  Y.~Gao, A.~V.~Gritsan, Z.~Guo, K.~Melnikov, M.~Schulze and N.~V.~Tran,
  Phys.\ Rev.\ D {\bf 81} (2010) 075022
  [arXiv:1001.3396 [hep-ph]], two lines after Eq.~(A2), and Eq.~(9) of
M.~C.~Kumar, P.~Mathews, A.~A.~Pankov, N.~Paver, V.~Ravindran and A.~V.~Tsytrinov,
  Phys.\ Rev.\ D {\bf 84} (2011) 115008
  [arXiv:1108.3764 [hep-ph]];
S.~Y.~Choi, D.~J.~Miller, M.~M.~Muhlleitner and P.~M.~Zerwas,
  Phys.\ Lett.\  B {\bf 553} (2003) 61
  [arXiv:hep-ph/0210077];
K.~Odagiri,
  JHEP {\bf 0303} (2003) 009
  [arXiv:hep-ph/0212215];
C.~P.~Buszello, I.~Fleck, P.~Marquard and J.~J.~van der Bij,
  Eur.\ Phys.\ J.\  C {\bf 32} (2004) 209
  [arXiv:hep-ph/0212396];
A.~Djouadi,
  Phys.\ Rept.\  {\bf 457} (2008) 1
  [arXiv:hep-ph/0503172];
  C.~P.~Buszello and P.~Marquard,
  arXiv:hep-ph/0603209;
A.~Bredenstein, A.~Denner, S.~Dittmaier and M.~M.~Weber,
  Phys.\ Rev.\  D {\bf 74} (2006) 013004
  [arXiv:hep-ph/0604011];
 P.~S.~Bhupal Dev, A.~Djouadi, R.~M.~Godbole, M.~M.~Muhlleitner and S.~D.~Rindani,
  Phys.\ Rev.\ Lett.\  {\bf 100} (2008) 051801
  [arXiv:0707.2878 [hep-ph]];
R.~M.~Godbole, D.~J.~Miller and M.~M.~Muhlleitner,
  JHEP {\bf 0712} (2007) 031
  [arXiv:0708.0458 [hep-ph]];
K.~Hagiwara, Q.~Li and K.~Mawatari,
  JHEP {\bf 0907} (2009) 101
  [arXiv:0905.4314 [hep-ph]];
A.~De Rujula, J.~Lykken, M.~Pierini, C.~Rogan and M.~Spiropulu,
  Phys.\ Rev.\  D {\bf 82} (2010) 013003
  [arXiv:1001.5300 [hep-ph]];
 C.~Englert, C.~Hackstein and M.~Spannowsky,
  Phys.\ Rev.\  D {\bf 82} (2010) 114024
  [arXiv:1010.0676 [hep-ph]];
  U.~De Sanctis, M.~Fabbrichesi and A.~Tonero,
  Phys.\ Rev.\  D {\bf 84} (2011) 015013
  [arXiv:1103.1973 [hep-ph]];
  V.~Barger and P.~Huang,
  Phys.\ Rev.\ D {\bf 84} (2011) 093001
  [arXiv:1107.4131 [hep-ph]];
  S.~Bolognesi, Y.~Gao, A.~V.~Gritsan, K.~Melnikov, M.~Schulze, N.~V.~Tran and A.~Whitbeck,
  arXiv:1208.4018 [hep-ph];
  R.~Boughezal, T.~J.~LeCompte and F.~Petriello,
  arXiv:1208.4311 [hep-ph];
  D.~Stolarski and R.~Vega-Morales,
  arXiv:1208.4840 [hep-ph];
S.~Y.~Choi, M.~M.~Muhlleitner and P.~M.~Zerwas,
arXiv:1209.5268 [hep-ph];
P.~Avery {\it et al.}; 
arXiv:1210.0896 [hep-ph];
C-Q.~Geng, D.~Huang, Y.~Tang and Y-L.~Wu,
arXiv:1210.5103 [hep-ph];
T.~Modak, D.~Sahoo and R.~Sinha,
arXiv:1301.5404 [hep-ph].

\bibitem{spin0}
S.~Chatrchyan {\it et al.}  [CMS Collaboration],
  Phys.\ Rev.\ Lett.\  {\bf 110} (2013) 081803
  [arXiv:1212.6639 [hep-ex]];
G.~Aad {\it et al.}  [ATLAS Collaboration],
  Phys.\ Lett.\ B {\bf 726} (2013) 120
  [arXiv:1307.1432 [hep-ex]].
  
  \bibitem{D0spin}
  D0 Collaboration,
  D0 Note 6387-CONF, {\tt http://www-d0.fnal.gov/Run2Physics/WWW/results/prelim/HIGGS/H138/H138.pdf}.
  
  \bibitem{eft}
  W.~Buchmuller and D.~Wyler,
  Nucl.\ Phys.\ B {\bf 268} (1986) 621.
  K.~Hagiwara, S.~Ishihara, R.~Szalapski and D.~Zeppenfeld,
  Phys.\ Rev.\ D {\bf 48}, 2182 (1993).
   K.~Hagiwara, R.~Szalapski and D.~Zeppenfeld,
  Phys.\ Lett.\ B {\bf 318}, 155 (1993)
  [hep-ph/9308347].
  B.~Grzadkowski, M.~Iskrzynski, M.~Misiak and J.~Rosiek,
  JHEP {\bf 1010}, 085 (2010)
  [arXiv:1008.4884 [hep-ph]].
  M.~B.~Einhorn and J.~Wudka,
  Nucl.\ Phys.\ B {\bf 876} (2013) 556
  [arXiv:1307.0478 [hep-ph]].
  
  \bibitem{EFTreview}
   S.~Willenbrock and C.~Zhang,
  arXiv:1401.0470 [hep-ph].
  
  \bibitem{dim6subset}
  F.~Bonnet, M.~B.~Gavela, T.~Ota and W.~Winter,
  Phys.\ Rev.\ D {\bf 85} (2012) 035016
  [arXiv:1105.5140 [hep-ph]].
   T.~Corbett, O.~J.~P.~Eboli, J.~Gonzalez-Fraile and M.~C.~Gonzalez-Garcia,
  arXiv:1207.1344 [hep-ph].
  W.~-F.~Chang, W.~-P.~Pan and F.~Xu,
  Phys.\ Rev.\ D {\bf 88} (2013) 3,  033004
  [arXiv:1303.7035 [hep-ph]].
  A.~Hayreter and G.~Valencia,
  Phys.\ Rev.\ D {\bf 88} (2013) 034033
  [arXiv:1304.6976 [hep-ph]].
  J.~Elias-Miro, J.~R.~Espinosa, E.~Masso and A.~Pomarol,
  JHEP {\bf 1311} (2013) 066
  [arXiv:1308.1879 [hep-ph]].
  S.~Banerjee, S.~Mukhopadhyay and B.~Mukhopadhyaya,
  Phys.\ Rev.\ D {\bf 89} (2014) 053010
  [arXiv:1308.4860 [hep-ph]].
  E.~Boos, V.~Bunichev, M.~Dubinin and Y.~Kurihara,
  Phys.\ Rev.\ D {\bf 89} (2014) 035001
  [arXiv:1309.5410 [hep-ph]].
  M.~Dahiya, S.~Dutta and R.~Islam,
  arXiv:1311.4523 [hep-ph].
  J.~S.~Gainer, J.~Lykken, K.~T.~Matchev, S.~Mrenna and M.~Park,
  arXiv:1403.4951 [hep-ph].
  
  
   \bibitem{eduard}
  E.~Masso and V.~Sanz, 
  Phys.\ Rev.\ D {\bf 87} (2013) 3,  033001
  [arXiv:1211.1320 [hep-ph]].
  
  \bibitem{hanandskiba}
  Z.~Han and W.~Skiba,
  Phys.\ Rev.\ D {\bf 71}, 075009 (2005)
  [hep-ph/0412166].
  
  \bibitem{dim6global}
  T.~Corbett, O.~J.~P.~Ebol, J.~Gonzalez-Fraile and M.~C.~Gonzalez-Garcia,
arXiv:1211.4580 [hep-ph];
   B.~Dumont, S.~Fichet and G.~von Gersdorff,
  JHEP {\bf 1307}, 065 (2013)
  [arXiv:1304.3369 [hep-ph]].
  
   \bibitem{pomarolriva}
  A.~Pomarol and F.~Riva,
  JHEP {\bf 1401}, 151 (2014)
  [arXiv:1308.2803 [hep-ph]].
  
  
  \bibitem{ewpt}
  S.~Alam, S.~Dawson and R.~Szalapski,
  Phys.\ Rev.\ D {\bf 57}, 1577 (1998)
  [hep-ph/9706542].
   A.~De Rujula, M.~B.~Gavela, P.~Hernandez and E.~Masso,
  Nucl.\ Phys.\ B {\bf 384}, 3 (1992).
   H.~Mebane, N.~Greiner, C.~Zhang and S.~Willenbrock,
  Phys.\ Rev.\ D {\bf 88}, no. 1, 015028 (2013)
  [arXiv:1306.3380 [hep-ph]].
  
 \bibitem{tgc}
    T.~Corbett, O.~J.~P.~Eboli, J.~Gonzalez-Fraile and M.~C.~Gonzalez-Garcia,
  Phys.\ Rev.\ Lett.\  {\bf 111}, no. 1, 011801 (2013)
  [arXiv:1304.1151 [hep-ph]].
  
  
  \bibitem{leshouches}
Contribution to the Les Houches 2013 Proceedings, {\it Triple gauge couplings
revisited}, A.~Falkowski, S.~Fichet, K.~Mohan, F.~Riva and V.~Sanz, to appear.

\bibitem{RGEbounds}
C.~Grojean, E.~E.~Jenkins, A.~V.~Manohar and M.~Trott,
  JHEP {\bf 1304} (2013) 016
  [arXiv:1301.2588 [hep-ph]].
  J.~Elias-Miró, J.~R.~Espinosa, E.~Masso and A.~Pomarol,
  JHEP {\bf 1308} (2013) 033
  [arXiv:1302.5661 [hep-ph]].
  J.~Elias-Miro, J.~R.~Espinosa, E.~Masso and A.~Pomarol,
  JHEP {\bf 1311} (2013) 066
  [arXiv:1308.1879 [hep-ph]].
  E.~E.~Jenkins, A.~V.~Manohar and M.~Trott,
  JHEP {\bf 1310} (2013) 087
  [arXiv:1308.2627 [hep-ph]].
  E.~E.~Jenkins, A.~V.~Manohar and M.~Trott,
  JHEP {\bf 1401} (2014) 035
  [arXiv:1310.4838 [hep-ph]].
  R.~Alonso, E.~E.~Jenkins, A.~V.~Manohar and M.~Trott,
  arXiv:1312.2014 [hep-ph].
    J.~Elias-Miro, C.~Grojean, R.~S.~Gupta and D.~Marzocca,
  arXiv:1312.2928 [hep-ph].
  
 \bibitem{dawson}
 C.~-Y.~Chen, S.~Dawson and C.~Zhang,
  Phys.\ Rev.\ D {\bf 89}, 015016 (2014)
  [arXiv:1311.3107 [hep-ph]].
   H.~Mebane, N.~Greiner, C.~Zhang and S.~Willenbrock,
  Phys.\ Rev.\ D {\bf 88}, no. 1, 015028 (2013)
  [arXiv:1306.3380 [hep-ph]].
  
  \bibitem{precisionhiggs}
  B.~Henning, X.~Lu and H.~Murayama,
  arXiv:1404.1058 [hep-ph].
  
\bibitem{D0comb}
 V.~M.~Abazov {\it et al.}  [D0 Collaboration],
  Phys.\ Rev.\ Lett.\  {\bf 109}, 121802 (2012)
  [arXiv:1207.6631 [hep-ex]].
  
\bibitem{atlasptv}
  ATLAS Collaboration,
  ATLAS-CONF-2013-079, {\tt https://cds.cern.ch/record/1563235/files/ATLAS-CONF-2013-079.pdf}.

\bibitem{ESY2}
 J.~Ellis, V.~Sanz and T.~You,
  Eur.\ Phys.\ J.\ C {\bf 73} (2013) 2507
  [arXiv:1303.0208 [hep-ph]].
  G.~Isidori and M.~Trott,
  JHEP {\bf 1402}, 082 (2014)
  [arXiv:1307.4051 [hep-ph], arXiv:1307.4051].
  
 \bibitem{silh}
 G.~F.~Giudice, C.~Grojean, A.~Pomarol and R.~Rattazzi,
  JHEP {\bf 0706}, 045 (2007)
  [hep-ph/0703164].
R.~Contino, C.~Grojean, M.~Moretti, F.~Piccinini and R.~Rattazzi,
  JHEP {\bf 1005} (2010) 089
  [arXiv:1002.1011 [hep-ph]];
  R.~Contino,
arXiv:1005.4269 [hep-ph];
 R.~Grober and M.~Muhlleitner,
  JHEP {\bf 1106} (2011) 020
  [arXiv:1012.1562 [hep-ph]].


\bibitem{benj}
  A.~Alloul, B.~Fuks and V.~Sanz,
  arXiv:1310.5150 [hep-ph], to be published in JHEP.
 
 \bibitem{mg5}
 J.~Alwall, M.~Herquet, F.~Maltoni, O.~Mattelaer and T.~Stelzer,
  JHEP {\bf 1106}, 128 (2011)
  [arXiv:1106.0522 [hep-ph]].
  
 \bibitem{pythia}
  T.~Sjostrand, S.~Mrenna and P.~Z.~Skands,
  JHEP {\bf 0605}, 026 (2006)
  [hep-ph/0603175].
  
  
 \bibitem{delphes}
   J.~de Favereau {\it et al.}  [DELPHES 3 Collaboration],
  JHEP {\bf 1402}, 057 (2014)
  [arXiv:1307.6346 [hep-ex]].
  
  \bibitem{fastrack}
  J.~Ellis, D.~S.~Hwang, V.~Sanz and T.~You,
  JHEP {\bf 1211} (2012) 134
  [arXiv:1208.6002 [hep-ph]].
  
  \bibitem{ciaran}
  V.~Sanz and C.~Williams, in preparation.
  
  
  
  \bibitem{MCFM}
    J.~M.~Campbell and R.~K.~Ellis,
  Nucl.\ Phys.\ Proc.\ Suppl.\  {\bf 205-206}, 10 (2010)
  [arXiv:1007.3492 [hep-ph]];
    R.~K.~Ellis,
  Nucl.\ Phys.\ Proc.\ Suppl.\  {\bf 160}, 170 (2006).
  J.~M.~Campbell,
  hep-ph/0105226.
   J.~M.~Campbell, R.~K.~Ellis and C.~Williams,
  JHEP {\bf 1107}, 018 (2011)
  [arXiv:1105.0020 [hep-ph]].
  
 \bibitem{silh-maggie} 
 R.~Contino, M.~Ghezzi, C.~Grojean, M.~Muhlleitner and M.~Spira,
  JHEP {\bf 1307}, 035 (2013)
  [arXiv:1303.3876 [hep-ph]].
 R.~Contino, M.~Ghezzi, C.~Grojean, M.~Muhlleitner and M.~Spira,
  arXiv:1403.3381 [hep-ph].
    
  \bibitem{presentationLHCresults}
  F.~Boudjema, G.~Cacciapaglia, K.~Cranmer, G.~Dissertori, A.~Deandrea, G.~Drieu la Rochelle, B.~Dumont and U.~Ellwanger {\it et al.},
  arXiv:1307.5865 [hep-ph].
  
  \bibitem{ATLASlikelihoods}
  G.~Aad {\it et al.}  [ATLAS Collaboration],
  Phys.\ Lett.\ B {\bf 726} (2013) 88
  [arXiv:1307.1427 [hep-ex]].
  
  \bibitem{ATLAStautau}
  ATLAS Collaboration, 
  ATLAS-CONF-2013-034 {\tt https://cds.cern.ch/record/1528170/files/ATLAS-CONF-2013-034.pdf}.
  
  \bibitem{CMSlikelihoods}
  CMS Collaboration,  
  CMS-PAS-HIG-13-005 {\tt http://cds.cern.ch/record/1542387/files/HIG-13-005-pas.pdf}.
  
  \bibitem{Contino}
    A.~Azatov, R.~Contino and J.~Galloway,
  JHEP {\bf 1204}, 127 (2012)
  [Erratum-ibid.\  {\bf 1304}, 140 (2013)]
  [arXiv:1202.3415 [hep-ph]].
  
  \bibitem{D02lepton}
  V.~M.~Abazov {\it et al.}  [D0 Collaboration],
  Phys.\ Rev.\ D {\bf 88} (2013) 052010
  [arXiv:1303.3276 [hep-ex]].
  
  \bibitem{D0jet}
  V.~M.~Abazov {\it et al.}  [D0 Collaboration],
  arXiv:1312.7623 [hep-ex].
  
    \bibitem{D01lepton}
  V.~M.~Abazov {\it et al.}  [D0 Collaboration],
  Phys.\ Rev.\ D {\bf 88} (2013) 052008
  [arXiv:1301.6122 [hep-ex]].
  
  \bibitem{D0zerolepton}
  V.~M.~Abazov {\it et al.}  [D0 Collaboration],
  Phys.\ Lett.\ B {\bf 716} (2012) 285
  [arXiv:1207.5689 [hep-ex]].


\bibitem{BKKLR}
  A.~Biekoetter, A.~Knochel, M.~Kraemer, D.~Liu and F.~Riva,
  arXiv:1406.7320 [hep-ph].
  
  

   
\end{thebibliography}
 \providecommand{\href}[2]{#2}\begingroup\raggedright\endgroup

\end{document}